# Astro2020 Science White Paper

# The Magellanic Stream as a Probe of Astrophysics

**Thematic Areas:** ☐ Planetary Systems ☐ Star and Planet Formation ☐ Formation and Evolution of Compact Objects ☐ Cosmology and Fundamental Physics ☐ Stars and Stellar Evolution ☐ Resolved Stellar Populations and their Environments ☒ Galaxy Evolution ☐ Multi-Messenger Astronomy and Astrophysics

**Principal Author:**
**Andrew J Fox**, Space Telescope Science Institute, afox@stsci.edu, 410 338 5083

**Co-authors:**
**Kathleen A. Barger,** Texas Christian University, k.barger@tcu.edu
**Joss Bland-Hawthorn,** University of Sydney, jbh@phys.usyd.edu.au
**Dana Casetti-Dinescu,** Southern Connecticut State University, casettid1@southernct.edu
**Elena d'Onghia,** University of Wisconsin-Madison, edonghia@astro.wisc.edu
**Felix J. Lockman**, Green Bank Observatory, jlockman@nrao.edu
**Naomi McClure-Griffiths**, Australian National University, naomi.mcclure-griffiths@anu.edu.au
**David Nidever,** University of Montana, david.nidever@montana.edu
**Mary Putman,** Columbia University, mputman@astro.columbia.edu
**Philipp Richter,** University of Potsdam, prichter@astro.physik.uni-potsdam.de
**Snezana Stanimirovic,** University of Wisconsin-Madison, sstanimi@astro.wisc.edu
**Thorsten Tepper-Garcia,** University of Sydney, tepper@physics.usyd.edu.au

**Abstract**:
Extending for over 200 degrees across the sky, the Magellanic Stream together with its Leading Arm is the most spectacular example of a gaseous stream in the local Universe. The Stream is an interwoven tail of filaments trailing the Magellanic Clouds as they orbit the Milky Way. Thought to be created by tidal forces, ram pressure, and halo interactions, the Stream is a benchmark for dynamical models of the Magellanic System, a case study for gas accretion and dwarf-galaxy accretion onto galaxies, a probe of the outer halo, and the bearer of more gas mass than all other Galactic high velocity clouds combined. If it survives to reach the Galactic disk, it may maintain or even elevate the Galactic star-formation rate. In this white paper, we emphasize the Stream's importance for many areas of Galactic astronomy, summarize key unanswered questions, and identify future observations and simulations needed to resolve them. We stress the importance of ultraviolet, optical, and radio spectroscopy, and the need for computational models that capture full particle and radiation treatments within an MHD environment.

# 1) Introduction

Since its discovery in 21 cm emission over 50 years ago (Dieter 1965, Wannier & Wrixon 1972, Mathewson+ 1974), the Magellanic Stream has fascinated many communities of astronomers, from radio and ultraviolet observers to dynamicists and simulators. At the simplest level, the Stream is an extended tail of multi-phase gas stripped out of the Magellanic Clouds and covering 140 degrees in length, or 200 degrees when including its Leading Arm (see Figure 1; Nidever+ 2008, 2010). Yet it is so much more than that: a case study of the accretion of gas and satellites onto a star-forming galaxy, a key constraint on the dynamical history of the Magellanic Clouds, a testbed for the evaporative encounters between cool gas clouds and the hot Galactic corona, a laboratory for understanding how star formation occurs in tidal tails, and a screen against which ionizing radiation from the Galactic Center shines (see review by D'Onghia & Fox 2016). For these diverse reasons, many sub-fields of Galactic astronomy are directly impacted by our understanding of the Stream.

Considerable progress on the Stream has been made over the last two decades, particularly via the use of ultraviolet (UV) spectrographs on *HST*, sensitive 21 cm radio surveys, and numerical simulations (Figure 2; D'Onghia & Fox 2016). However, many open questions remain, including fundamental properties like its distance, origin and fate. In this white paper, we identify progress made in the last few years (Section 2), and then outline remaining questions to be answered (Section 3). We then focus on future observational capabilities and necessary refinements to state-of-the-art simulations (Section 4), and we finish with some concluding remarks (Section 5). Throughout the white paper we emphasize the Stream's importance as a probe of astrophysical processes and the necessity of spectroscopic observations as diagnostic tools.

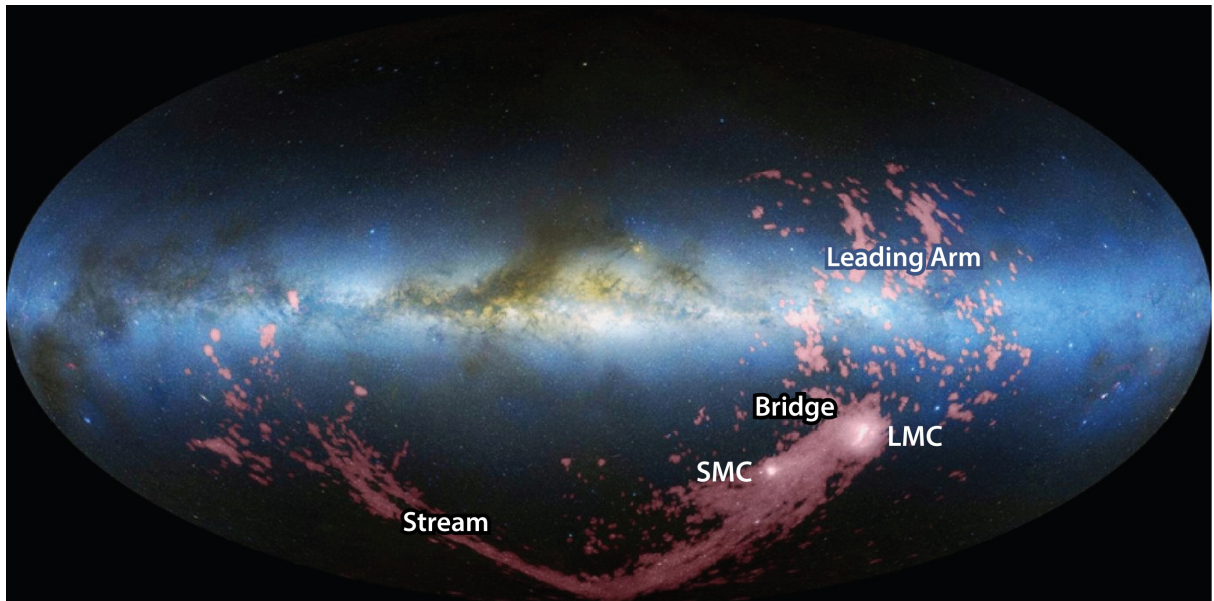

Figure 1: All-sky H I map of the Magellanic System, from Nidever+ (2010), using H I data from the GBT, Arecibo, Parkes, Westerbork, and the LAB survey. This Aitoff projection is in Galactic coordinates with H I emission colored in pink. We refer to the entire complex (LMC, SMC, Stream, Bridge, and Leading Arm) as the Magellanic System.

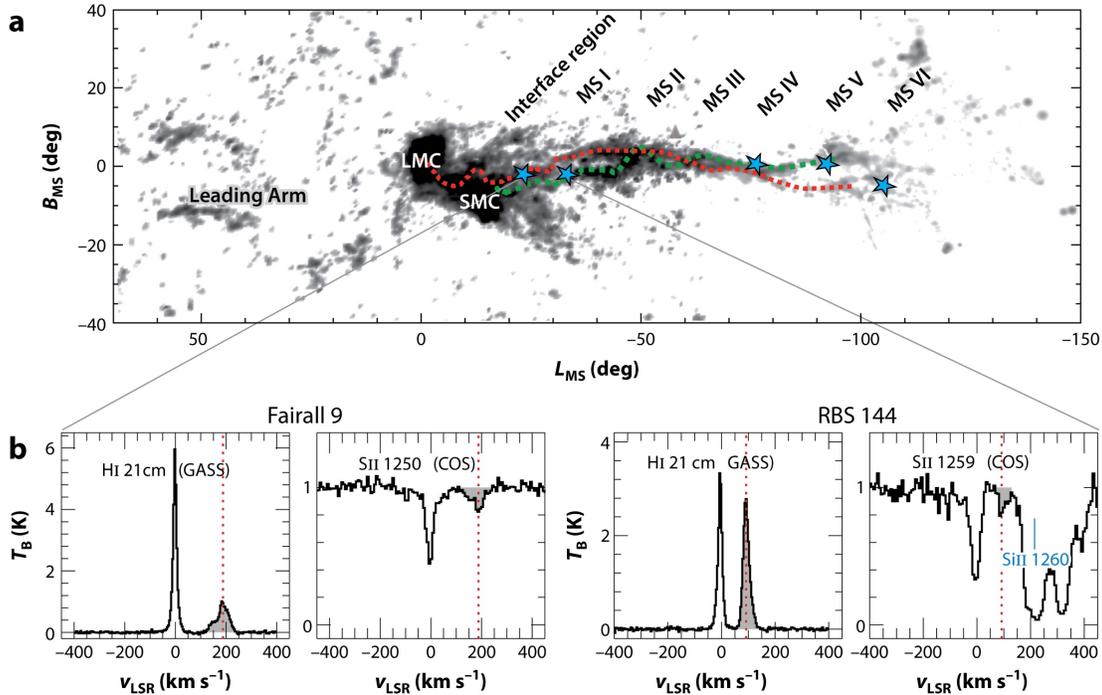

Figure 2: Illustration of the chemical abundances along the Magellanic Stream, from D'Onghia & Fox (2016), and using data from Fox+ (2013) and Richter+ (2013). The two main filaments of gas (Nidever+ 2008) have markedly different kinematic and chemical properties, indicating the Stream's dual nature, with LMC and SMC contributions.

## 2) Recent Progress on the Magellanic Stream and Magellanic System

a. **Over ten ultra-faint dwarf (UFD) galaxies** have been discovered in the vicinity of the Magellanic Clouds (Koposov+ 2015a,b, 2018, Bechtol+ 2015, Drlica-Wagner+ 2016), using data from the Dark Energy Survey (DES). The proximity of these satellites to the Magellanic Clouds suggests a *Magellanic group* of galaxies (e.g. Kallivayalil+ 2018, Fritz+ 2019), which may have influenced the formation of the Stream (Tepper-Garcia+ 2019).  More work is needed to identify which of the *bright* dwarf spheroidals have a Magellanic association (Lynden-Bell 1976, D'Onghia & Lake 2008, Nichols+ 2011). The LMC has a fairly massive companion, the SMC, which is 1.5 mag fainter, but its next most luminous satellite appears to be Hydrus 1 (Koposov+ 2018), nearly 13 mag fainter. This leaves an unexplained >10 mag gap in the LMC satellite luminosity function. Gaia DR2 proper motions suggest that two dwarf spheroidals have orbits closely aligned with the Magellanic-Cloud plane (Pardy+ 2019). Together with the SMC this raises to three the number of LMC satellites in the mass range of dwarf spheroidals, in agreement with $\Lambda$CDM predictions.

b. **The metallicity of the Leading Arm** has been constrained via UV spectroscopy from *HST*/COS (Fox+ 2018, Richter+ 2018a). These studies have found the Leading Arm has an SMC-like chemical abundance composition, but with considerable variation between the different regions. The oxygen abundances vary from 4% solar to 30% solar between regions LA II and LA III (Fox+ 2018), suggesting *multiple gas removal episodes*. While there is kinematic and chemical evidence for a filament in the *trailing Stream* that originates in the LMC (Nidever+ 2008, Richter+ 2013), as yet no chemical evidence for an LMC filament in the *Leading Arm*, though H I studies (Putman+ 1998) favor an LMC origin. Further UV studies from sensitive future space-based instrumentation could map out the chemical abundances in the Leading Arm and search for a LMC filament.

c. **Some short-length stellar components to the Stream** have been found in the DES data (Belokurov & Koposov 2016; Navarrete+ 2019), one of which overlaps with the gaseous stream, but we are missing a breakthrough discovery of an extended stellar stream (a Sagittarius dwarf). The absence of a stellar stream has been a long-standing problem, since a tidally-created Stream should contain both stars and gas. The Stream somehow survives yet does not form stars (Stanimirovic+ 2010). Other surveys are needed, such as the Survey of the Magellanic Stellar History (SMASH) using the Dark Energy Camera, which has mapped 480 square degrees of the outskirts of the Magellanic Clouds (Nidever+ 2017). LSST will be instrumental in this effort.

d. **Proper-motion and parallax measurements** for stars in the LMC and SMC and for the ultra-faint dwarfs are now possible with *Gaia DR2*. Ongoing *HST* proper-motion measurements will better constrain the LMC's and SMC's internal kinematics (e.g. Oey+ 2018), further constrain dynamical models of the entire Magellanic System, and explore evidence for a direct LMC-SMC collision in the past (Besla+ 2012). For the UFDs, systemic proper motions will determine whether they are Magellanic members. These measurements are still uncertain for some systems due to the small number of stars that have Gaia DR2 measurements (e.g. Kallivayalil+ 2018, Fritz+ 2019). An accurate census of Magellanic UFDs requires further improved proper motions. Specifically, precise proper motions of faint stars must be acquired, and for this purpose *JWST* is key.

e. **Stream-analogs in other galaxies** have been identified. For example, the Whale Galaxy (NGC 4631), shows evidence for a massive tidal gas stream with an oxygen abundance of 13% solar and an estimated total gas mass of $10^9$ solar masses (Richter+ 2018b), which closely matches the mass and metallicity of the Magellanic Stream (Fox+ 2014). Richter+ (2018b) conclude that the tidal stream in the Whale galaxy represents gas stripped from satellite galaxies. The nearby spirals M31 and M33 are connected by a bridge of H I (Braun & Thilker 2004, Wolfe+ 2013), although it is unclear whether it represents a condensing intergalactic filament rather than a tidal feature. Finally, some tidal tails are seen around interacting dwarfs (Pearson+ 2016).

f. **Small-scale structure in the Stream** has been studied via detailed H I studies (Kalberla & Haud 2006, Stanimirovic+ 2008, Matthews+ 2009, Nigra+ 2012, For+ 2014). Curiously, cold H I cores are seen yet heating/cooling equilibrium calculations predict that no cold neutral medium should exist beyond 25 kpc. This issue needs to be resolved with future high-resolution H I observations.

### 3) Remaining Questions for the Future

**f. Distance.** This key parameter has implications for the Stream's total mass and fate, and the orbital history of the Magellanic Clouds. Yet the only solid distance constraint on the Stream is that one end is anchored to the Magellanic Clouds, at 55 kpc. The distance to the Leading Arm is also poorly known, though some regions have constraints ranging from ~20-30 kpc (McClure-Griffiths+ 2008, Price-Whelan+ 2018). In a first-passage scenario, the Leading Arm should be approximately at the distance of the Clouds since they are currently near perigalacticon, so distance measurements can directly test this scenario. Upcoming *Gaia* data releases can be used to select stellar targets (e.g. blue horizontal branch stars) for optical absorption-line studies of the Stream using Ca II and Na I, to provide direct distance constraints on the Stream.

g. **Mass inflow rate**. What is the mass inflow rate in the Stream and how does it compare to the Galactic star formation rate? Current estimates of the Stream's inflow rate are in the range 3-7 solar masses per year (Fox+ 2014, Richter+ 2017), though they scale with the distance to the Stream, so may need to be revised upward. This will allow us to assess whether the Stream will trigger a future starburst in the Milky Way, depending on its interaction with the halo.

h. **Fate**. The fate of the Stream is still unknown. Whether the cool gas will eventually settle onto the Galactic disk depends on the nature of the interaction with the hot Galactic corona. Simulations have revealed that this interaction can be both evaporative (e.g. Hensler & Vieser 2002, Heitsch & Putman 2009) or condensative (Fraternali+ 2015, Armillotta+ 2017), so that cool clouds can either shrink or grow with time, depending on the metallicity and density contrast with the surrounding hot medium. Better constraints on the hot Galactic halo with future X-ray facilities will improve our understanding of the external medium. Furthermore, maps of the magnetic field across the Stream are needed, because magnetic fields may stabilize the cloud against ram-pressure stripping and conductive evaporation, and thus impact the Stream's fate. McClure-Griffiths+ (2010) reported a 6 $\mu$G coherent field in a Leading Arm cloud based on Faraday rotation measures, and Kaczmarek+ (2017) find a 0.3 $\mu$G coherent field in the Magellanic Bridge.

i. **Ionization by the Galactic Center**. Bland-Hawthorn+ (2013, 2019) have reported an imprint of a Seyfert flare at the Galactic Center in the Magellanic Stream. In this scenario, a flare several Myr ago released a burst of ionizing radiation that ionized the Stream, which is now recombining and glowing in H$\alpha$ emission. The Stream may thus probe *recent nuclear activity* from Sgr A*. Further UV absorption-line ratios (e.g. C IV/C II, Si IV/Si II; Fox+ 2014) and optical H$\alpha$ emission-line observations (Putman+ 2003, Barger+ 2017) can test this scenario, and explore whether its signature can be distinguished from other ionization processes in the halo, such as shocks (Bland-Hawthorn+ 2007, Tepper-Garcia+ 2015). Optical spectroscopic facilities (e.g. the Wisconsin H-Alpha Mapper) that can map emission from H$\alpha$ and other nebular lines ([S II], [N II], [O II]) across the entire Stream are needed to complete our multi-phase view of the Stream.

j. **Total spatial extent**. Further UV and radio surveys are needed to address the Stream's total footprint on the sky, both in neutral and ionized gas. Does its tip cross the Galactic plane? Do the Stream and Leading Arm together form a great circle? Current all-sky surveys (e.g. the HI4PI survey, HI4PI collaboration, 2018) reach column densities of a few x$10^{18}$ cm$^{-2}$; future facilities and surveys (GASKAP, FAST) reaching a few x$10^{17}$ cm$^{-2}$ would reveal considerably more structure, since the H I column density distribution function rises to low $N$(H I). Indeed, GBT observations of small regions of the Stream already reveal structure at a few x$10^{17}$ cm$^{-2}$ (Howk+ 2017). Charting the full size of the Stream will further constrain the dynamics of the entire Magellanic System, informing whether the Clouds are at their first passage around the MW (Besla+ 2007, 2010).

### 4) Future Progress

The UV, optical, and radio are the key regimes for diagnosing the Stream's physical and chemical conditions. In terms of spectral lines per Angstrom, the UV is the richest portion of the electromagnetic spectrum with a large number of ionization states available. The Stream's composition and physical properties can be determined from full analyses of these UV absorption lines in the spectra of background sources. High-resolution UV spectrographs with multiplexing capabilities on large-aperture space telescopes (such as *LUVOIR* or *HabEx*) would provide significantly higher sensitivity than the current UV spectrographs on *Hubble*, yielding an order of magnitude more background targets, allowing the Stream's chemical abundance patterns to be mapped out on finer spatial scales. We also need *optical spectroscopy* of any stars discovered toward the Stream and Leading Arm (Casetti-Dinescu+ 2014, Price-Whelan+ 2018, Zhang+ 2017, 2019) to verify their membership via kinematic and abundance analyses. Spectroscopy of the main sequence at the distance of the Stream demands large telescopes with multiplexing ability.

On the radio front, continued 21 cm studies are needed to reveal the full extent and kinematic structure of the Stream (McClure-Griffiths+ 2018). The GASKAP H I survey (Dickey+ 2013) using the Australian Square Kilometer Array Pathfinder (ASKAP) telescope will map the Stream at 30" resolution, compared to 16.2' with the HI4PI survey, the current state-of-the-art. Observations from the North are also needed. Progress will be made by the new generation of focal-plane array receivers on the GBT and Arecibo, which offer an order-of-magnitude increase in mapping speeds and sensitivities to $N$(H I) of a few x$10^{17}$. The 500m FAST telescope (China) will contribute in the next decade at 3.5' resolution. Polarization measurements (e.g. from the POSSUM project on ASKAP) will make progress on the magnetic properties of the Stream and Leading Arm.

Only by combining the UV and radio data with high-resolution hydrodynamical simulations on scales of individual clouds can we fully understand the fate of the Stream. Full particle and radiation treatments within an MHD environment are needed. Magnetic effects (Gronnøw+ 2017, 2018) and heating effects (Hensler & Vieser 2002) are both important in HVCs. Recent hydrodynamical models have improved our understanding of the LMC-SMC system (Pardy+ 2018) and included the effect of ram-pressure stripping (Bustard+ 2018, Tepper-Garcia+ 2019). Nonetheless, further refinements are needed, including sub-grid physics such as metal-mixing and non-equilibrium cooling. Better algorithms, codes, and software/hardware architectures are needed to make progress. The Stream will always be the test case for these models.

## 5) Concluding Remarks

Understanding the Galactic halo is no less complex than studying the Earth's atmosphere. The processes of accretion, outflows, and gas recycling circulate material between the disk and the halo, just as gas circulation is a key process in the terrestrial atmosphere. The Magellanic Stream affords a nearby benchmark for the study of these processes, allowing us to study gas physics and metal mixing in close detail. The exchange of gas occurring when satellites like the Magellanic Clouds approach centrals like the Milky Way (a.k.a. intergalactic metal transfer) may be a significant and potentially dominant mode of gas transfer between galaxies, as suggested by modern hydrodynamic simulations (Angles-Alcazar+ 2017, Hafen+ 2018). By characterizing the Stream's properties, *we directly constrain* intergalactic metal transfer and thereby address the fundamental question of how galaxies get their gas. Extragalactic studies of the CGM and the baryon cycle have the advantage of large statistical samples, but do not have the spatial resolution or multi-wavelength datasets available in the Milky Way and Local Group. For these reasons, we endorse continued multi-pronged studies of the Magellanic Stream and the gaseous halo of the Milky Way using UV, radio, and optical spectroscopy.

## 6) References

Angles-Alcazar, D.+ 2017, MNRAS, 470, 4698
Armillotta, L.+ 2017, MNRAS, 470, 114
Barger, K.+ 2017, ApJ, 851, 110
Bechtol, K.+ 2015, ApJ, 807, 50
Belokurov & Koposov 2016, MNRAS, 456, 602
Besla, G.+ 2007, ApJ, 668, 949
Besla, G.+ 2010, ApJ, 721, L97
Besla, G.+ 2012, MNRAS, 421, 2109
Bland-Hawthorn, J.+ 2007, ApJL, 670, L109
Bland-Hawthorn, J.+ 2013, ApJ, 778, 58
Bland-Hawthorn, J.+ 2019, ApJ, subm.
Braun, R. & Thilker, D. 2004, A&A, 417, 421
Bruns, C.+ 2005, A&A, 432, 45
Bustard, C.+ 2018, ApJ, 863, 49
Casetti-Dinescu, D.+ 2014, ApJ, 784, L37
Dickey+ 2013, PASA, 30, 0003
Dieter, N. 1965, AJ, 70, 552
D'Onghia, E. & Fox, A. 2016, ARAA,
D'Onghia, E. & Lake, G. 2008, ApJL, 686, L61
Drlica-Wagner A.+ 2016, ApJ, 833, L5
For, B.-Q.+ 2014, ApJ, 792, 43
Fox, A.+ 2013, ApJ, 772, 110
Fox, A.+ 2014, ApJ, 787, 147
Fox, A.+ 2018, ApJ, 854, 142
Fraternali, F.+ 2015, MNRAS, 447, L70
Fritz, T.+ 2019, A&A, in press (arXiv:1805.07350)
Gronnøw, A.+ 2017, ApJ, 845, 69
Gronnøw, A.+ 2018, ApJ, 865, 64
Hafen, Z.+ 2018, MNRAS, subm. (arXiv:1811.11753)
Hensler, G. & Vieser, W. 2002, Ap&SS, 265, 397
Heitsch, F. & Putman, M. 2009, ApJ, 698, 1485
HI4PI Collaboration 2016, A&A, 594, A116
Howk, J. + 2017, ApJ, 846, 141
Kaczmarek, J.+ 2017, MNRAS, 467, 1776
Kalberla, P. & Haud, U. 2006, A&A, 455, 481

Kallivayalil, N. 2018, ApJ, 867, 19
Koposov, S.+ 2015a, ApJ, 805, 130
Koposov, S.+ 2015b, ApJ, 811, 62
Koposov, S.+ 2018, MNRAS, 479, 5343
Lynden-Bell, D. 1976 MNRAS 174:695
Mathewson+ 1974, ApJ, 190, 291
Matthews, D. 2009, ApJL, 691, L115
McClure-Griffiths, N.+ 2008, ApJL, 673, L143
McClure-Griffiths, N.+ 2010, ApJ, 725, 275
McClure-Griffiths, N.+ 2018, Nat. Astronomy, 2, 901
Nichols, M.+ 2011, ApJ, 742, 110
Navarrete, C.+ 2019, MNRAS, 483, 4160
Nidever, D.+ 2008, ApJ, 679, 432
Nidever, D.+ 2010, ApJ, 723, 1618
Nidever, D.+ 2017, AJ, 154, 199
Nigra, L. + 2012, ApJ, 760, 48
Oey, M.+ 2018, ApJ, 867, L8
Pardy, S.+ 2018, ApJ, 857, 101
Pardy, S.+ 2019, ApJ, under review
Pearson, S.+ 2016, MNRAS, 459, 1827
Price-Whelan, A.+ 2018, ApJ, sub. (arXiv:1811.05911)
Putman, M.+ 1998, Nature, 394, 752
Putman, M.+ 2003, ApJ, 597, 948
Richter, P.+ 2013, ApJ, 772, 111
Richter, P.+ 2018a, ApJ, 865, 145
Richter, P.+ 2018b, ApJ, 868, 112
Stanimirovic, S.+ 2008, ApJ, 680, 276
Stanimirovic, S.+ 2010, SerAJ, 180, 1
Tepper-Garcia+ 2015, ApJ, 813, 94
Tepper-Garcia+ 2019, MNRAS, sub. arXiv:1901.05636
Wannier, P. & Wrixon, G. 1972, ApJL, 173, L119
Wolfe, S.+ 2013, Nature, 497, 224
Zhang, L.+ 2017, ApJ, 835, 285
Zhang, L.+ 2019, ApJ, 871, 99